\pdfoutput=1
\documentclass[reprint,%
aps,prl,%
10pt,%
letterpaper,
superscriptaddress,%
amsmath,amssymb,%
floatfix,
longbibliography]{revtex4-2}

\usepackage{blindtext}
\usepackage{amssymb}
\usepackage{amsmath}
\usepackage{bm}
\usepackage{physics}

\usepackage{mathabx}

\usepackage{graphicx}
\graphicspath{ {./imgs_pdf/} }

\begin{document}

\title{Experimental Realization of Entangled Coherent States in Two-dimensional Harmonic Oscillators of a Trapped Ion}
\author{Honggi Jeon}
\affiliation{Department of Physics and Astronomy, Seoul National University, Seoul 08826, Republic of Korea}
\affiliation{Automation and System Research Institute, Seoul National University, Seoul 08826, Republic of Korea}
\author{Jiyong Kang}
\affiliation{Automation and System Research Institute, Seoul National University, Seoul 08826, Republic of Korea}
\affiliation{Department of Computer Science and Engineering, Seoul National University, Seoul 08826, Republic of Korea}
\author{Jaeun Kim}
\affiliation{Automation and System Research Institute, Seoul National University, Seoul 08826, Republic of Korea}
\affiliation{Department of Computer Science and Engineering, Seoul National University, Seoul 08826, Republic of Korea}
\author{Wonhyeong Choi}
\affiliation{Automation and System Research Institute, Seoul National University, Seoul 08826, Republic of Korea}
\affiliation{Department of Computer Science and Engineering, Seoul National University, Seoul 08826, Republic of Korea}
\affiliation{Inter-university Semiconductor Research Center, Seoul National University, Seoul 08826, Republic of Korea}
\author{Kyunghye Kim}
\affiliation{Automation and System Research Institute, Seoul National University, Seoul 08826, Republic of Korea}
\affiliation{Department of Computer Science and Engineering, Seoul National University, Seoul 08826, Republic of Korea}
\author{Taehyun Kim}
\email{taehyun@snu.ac.kr}
\affiliation{Automation and System Research Institute, Seoul National University, Seoul 08826, Republic of Korea}
\affiliation{Department of Computer Science and Engineering, Seoul National University, Seoul 08826, Republic of Korea}
\affiliation{Inter-university Semiconductor Research Center, Seoul National University, Seoul 08826, Republic of Korea}
\affiliation{Institute of Computer Technology, Seoul National University, Seoul 08826, Republic of Korea}
\affiliation{Institute of Applied Physics, Seoul National University, Seoul 08826, Republic of Korea}

\begin{abstract}
Entangled coherent states play pivotal roles in various fields such as quantum computation, quantum communication, and quantum sensing. We experimentally demonstrate the generation of entangled coherent states with the two-dimensional motion of a trapped ion system. Using Raman transitions with appropriate detunings, we simultaneously drive the red and blue sidebands of the two transverse axes of a single trapped ion and observe multi-periodic entanglement and disentanglement of its spin and two-dimensional motion. Then, by measuring the spin state, we herald entangled coherent states of the transverse motions of the trapped ion and observe the corresponding modulation in the parity of the phonon distribution of one of the harmonic oscillators. Lastly, we trap two ions in a linear chain and realize Mølmer–Sørensen gate using two-dimensional motion.
\end{abstract}

\maketitle
\section{INTRODUCTION}
For the last few decades, the coherent state has been the subject of intense theoretical and experimental investigation \cite{glauber_coherent_1963}. It is considered to be a quantum state with the most classical properties because its spread is the minimal allowed by the uncertainty principle and its trajectory of time evolution is identical to that of the classical harmonic oscillator \cite{sanders_review_2012}. Its multipartite extension, the entangled coherent state, has been a useful theoretical tool in various fields of quantum optics as it is the entangled superposition of the most "classical" quantum states. It has been used in theoretical studies concerning quantum information processing \cite{jeong_quantum-information_2001, jeong_efficient_2002, wang_quantum_2001, jeong_greenberger-horne-zeilinger--type_2006, park_entangled_2010, lee_near-deterministic_2013, van_enk_entangled_2001}, quantum metrology \cite{joo_quantum_2012}, and fundamental tests of physics such as Bell’s inequality and Leggett’s inequality \cite{stobinska_violation_2007, lee_faithful_2011}. Despite their sensitivity to decoherence, entangled coherent states have been experimentally realized in a few experiments involving photons \cite{ourjoumtsev_preparation_2009} and superconducting circuits \cite{wang_flying_2022, wang_schrodinger_2016}. The trapped ion has been an extremely valuable tool for studying the quantum world because it is highly isolated from the environment yet can be precisely controlled. The single-mode superposition of coherent states or cat states have been realized in trapped ion systems in various experiments using the motional state of the trapped ion \cite{monroe_schrodinger_1996, haljan_spin-dependent_2005, mcdonnell_long-lived_2007, poschinger_observing_2010, kienzler_observation_2016}. There have been several theoretical works on the implementation of entangled coherent states in trapped ion systems \cite{gerry_generation_1997, zou_generation_2001, solano_entangled_2002, zhong_generation_2018}, but none have been experimentally implemented so far.\\
\indent In this work, we report on the realization of entangled coherent states with the two dimensional motion of a trapped ion. We implement the simultaneous spin-dependent force (SDF) on the ion in the two principal axes (X and Y) by making the transverse trap potential nearly isotropic so that the secular frequencies of the X and Y modes are very close. By choosing a laser detuning between the X and Y mode frequencies and driving a bichromatic transition with the blue and red sidebands, we excite the motional modes in the two radial directions concurrently with varying ratios of coupling strengths to each mode. 
\par For a single ion, we generate Lissajous-curve-like motion in two dimensions with various commensurate oscillation periods in each direction and observe corresponding periodic variation in the spin state \cite{rossetti_trapped-ion_2016}. With mid-circuit measurement, we decouple the spin from the motion and herald the entangled coherent state of motion in two transverse axes. This is verified by observing the modulation of phonon number parity, which results from the periodic entanglement and disentanglement of the two motional modes. Also, in an ion chain consisting of two ions, we demonstrate the successful generation of a Bell state using Mølmer–Sørensen interaction where the geometric phase is accumulated via motion in two spatial dimensions, which reduces the required Rabi frequency compared to the one-dimension case.

\section{RESULTS}
\begin{figure*}[t]
  \includegraphics[width= \textwidth]{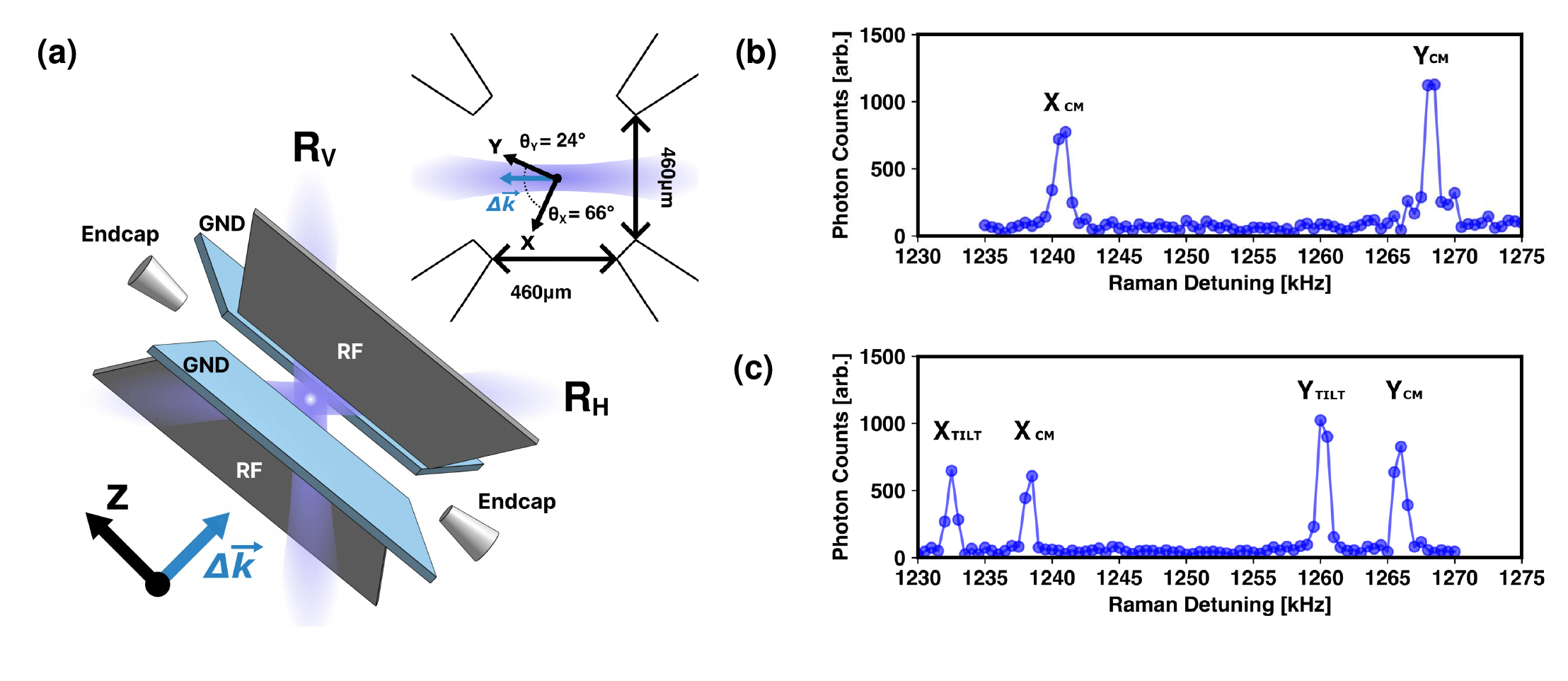}
  \centering
    \caption{\textbf{Experimental setup.} (a) Schematic diagram of trap electrodes, ion, and pulsed laser beams for Raman transition. $\Delta\vec{k}$ indicates the direction of momentum transfer, which is the difference of the two pulsed laser beams, $R_V$ and $R_H$. Upper right is the cross-section of the trap in the transverse plane. The angles between $\Delta\vec{k}$ and the Y and X principal axes are $24^\circ$ and $66^\circ$, respectively. The ion chain is formed along the Z axis. (b) Representative spectra showing the blue sidebands for the transverse modes of a single ion and (c) of a linear chain of two ions where $X_{cm}$ and $Y_{cm}$ are in-phase modes and $X_{tilt}$ and $Y_{tilt}$ are out-of-phase modes.}
\label{fig:trap}
\end{figure*}

We trap ${}^{171} \mathrm{Yb^{+}}$ ions in the center of a blade-type trap with dimensions specified in \ref{fig:trap}(a) \cite{jeon_coherent_2021, schmidt-kaler_how_2003}. A radio-frequency (RF) voltage oscillating at 15.3 MHz is applied to the RF blades for radial confinement, and the other two blades are grounded. A DC offset voltage can be applied to the RF blades to rotate the radial principal axes and change the separation between the X and Y mode secular frequencies. The amplitude of the RF field is actively stabilized by a PID controller \cite{johnson_active_2016}. As can be seen in \ref{fig:trap}(b) and (c), the mean secular frequency for the radial modes is typically set to around $\omega_{X,Y}\simeq2\pi\times1250$ kHz. The exact value changes by a few kHz every day due to the thermal drift of the RF voltage sampling  \cite{park_new_2021}. The degeneracy of the transverse axes is lifted by $2\pi\times27.8$ kHz, with the Y axis frequency higher. A DC voltage of 1300 V is applied to the endcap electrodes, and the resulting axial secular frequency is $\omega_{Z}=2\pi\times120\pm0.12$ kHz. The qubit states are defined as $\ket{\downarrow}=\ket{S_{1/2}, F=0, m_{F}=0}$ and $\ket{\uparrow}=\ket{S_{1/2}, F=1, m_{F}=0}$. For experiments involving a single ion, the qubit state is measured by the standard fluorescence detection method \cite{olmschenk_manipulation_2007}. For a two-ion chain, we use histogram fitting to infer the population of the three possible classes of qubit states, \{$\ket{\downdownarrows}$\}, \{$\ket{\downuparrows}$⟩, $\ket{\updownarrows}$\} and \{$\ket{\upuparrows}$\} \cite{turchette_deterministic_1998}. \\
\indent The red sideband, blue sideband, and carrier transitions are implemented by applying appropriate detunings to the stimulated Raman transition \cite{hayes_entanglement_2010}. It is realized by two perpendicular 355-nm pulsed laser beams which enter the trap from the bottom $(R_{V})$ and the side $(R_{H})$. Their relative frequencies and phases are controlled by acousto-optic modulators (AOMs). The Raman transition momentum vector $\Delta\vec{k}$ is perpendicular to the Z axis and has components in both radial axes with angles $\theta_{Y}=24^{\degree}$ and $\theta_{X}=66^{\degree}$  as shown in \ref{fig:trap}(a). This results in asymmetric Lamb-Dicke factors, $\eta_X = 0.05$ and $\eta_Y = 0.11$, for the modes. The beating of the pulsed laser is stabilized by a feed-forward system which shifts the driving RF frequency of the AOM that controls the $(R_{V})$ beam \cite{mount_scalable_2016}.

\subsection{Entanglement of spin with multiple motional modes}
\begin{figure*}[ht!]
  \centering
  \includegraphics[width= \textwidth]{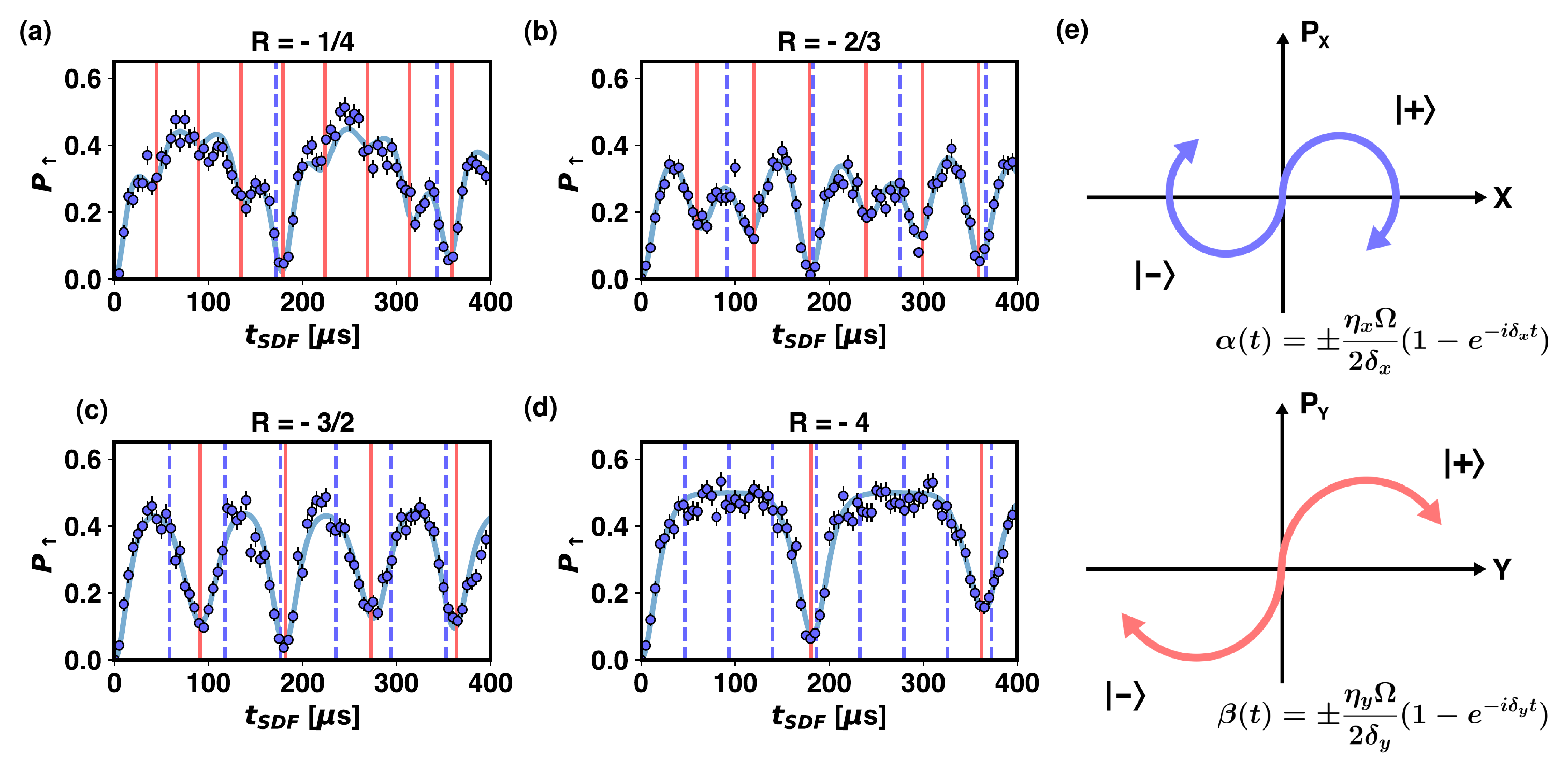}
    \caption{\textbf{Entanglement of spin and the two motional modes.} In (a-d) time evolution of the spin state for various detuning ratios is observed. Partial disentanglement of spin and motion takes place when wave packets return to the origin in only one dimension. Complete disentanglement is observed when wave packets return to the origin in both dimensions. Error bars indicate quantum projection noise. Solid curves are fits to \ref{eq3}. $R$ is the ratio of detunings to the radial modes, defined as $R=\delta_X/\delta_Y$. Values of $R$ estimated from fitting are (a) $-0.261\pm0.001$, (b) $-0.653\pm0.003$, (c) $-1.547\pm0.016$, and (d) $-3.892\pm0.071$. Times at which each motional mode is disentangled from the spin are indicated by vertical lines. Solid red lines correspond to the Y mode and dashed blue lines to the X mode. (e) A representative phase space diagram for the motional modes. In each phase space, the wave function evolves into a coherent superposition of two wave packets, corresponding to the $\phi_S$ basis spin eigenstates. The trajectories are determined by Rabi frequency and detuning from each mode.}
\label{fig:sdf}
\end{figure*}

We will use the notation $\ket{s}\ket{a}\ket{b}$ to specify the quantum state of the system, where $s$ denotes the qubit state of the ion chain with possible values of $\uparrow$ and $\downarrow$ for a single ion and their tensor product for a chain of two ions. $a$ and $b$ indicate the quantum states of the X and Y modes either in Fock state or coherent state basis. We realize the SDF Hamiltonian by driving the red and blue sidebands of the motional modes simultaneously with the same strength. When there is a symmetric detuning from the sidebands, the position of the wave packet in phase space modulated by a frequency proportional to the detuning \cite{haljan_spin-dependent_2005}, resulting in a circular trajectory as shown in \ref{fig:sdf}(e).
\par For a single trapped ion in a two-dimensional harmonic potential subject to a symmetrically detuned bichromatic beam, we have the following interaction Hamiltonian
\begin{equation}
\begin{split}
    \hat{H}=\frac{\hbar\mathrm{\Omega}\eta_{X}}{2}\left({\hat{a}}_{X}e^{-i\left(\delta_{X}t+\phi_{M}\right)}+{\hat{a}}_{X}^\dag e^{i\left(\delta_{X}t+\phi_{M}\right)}\right){\hat{\sigma}}_{\phi_{S}}+\\
    \frac{\hbar\mathrm{\Omega}\eta_Y}{2}\left({\hat{a}}_Ye^{-i(\delta_Yt+\phi_M)}+{\hat{a}}_Y^\dag e^{i(\delta_Yt+\phi_M)}\right){\hat{\sigma}}_{\phi_S}
\end{split}
\label{eq1}
\end{equation}
where $\eta_j$ and $\delta_j$ are the Lamb-Dicke factor for the $j$-th axis and the detuning from the center-of-mass mode of the $j$-th axis, respectively. $\hat{a_j}(\hat{a_j}^\dag)$ is the phonon annihilation (creation) operator for the $j$-th axis and $\Omega$ is the Rabi frequency of the Raman transition. The motion and spin phase of spin-dependent interaction is proportional to the difference, $\phi_M=(\phi_b-\phi_r)/2$, and sum, $\phi_S=(\phi_b+\phi_r)/2$, of the laser phases for the blue and red sidebands, $\phi_b$ and $\phi_r$. 

\par We will use $\ket{+}=1/\sqrt{2}(\ket{\downarrow}+e^{i\phi_S}\ket{\uparrow})$ and $\ket{-}=1/\sqrt{2}(\ket{\downarrow}-e^{-i\phi_S}\ket{\uparrow})$ to indicate the eigenstates of the ${\hat{\sigma}}_{\phi_S}$ operator. We assume  $\phi_S=0$ for simplicity. The X axis terms and Y axis terms act on their respective Hilbert spaces, yielding the following time evolution operator with displacements in X and Y phase spaces defined as $\alpha\left(t\right)=\eta_X\Omega/2\delta_X\left(1-e^{-i\delta_Xt}\right)e^{-i\phi_M}$ and $\beta\left(t\right)=\eta_Y\Omega/2\delta_Y\left(1-e^{-i\delta_Yt}\right)e^{-i\phi_M}$. The time evolution operator is $\hat{U}(t)=\ket{+}\bra{+}\hat{D}_X(\alpha(t))\hat{D}_Y(\beta(t))+\ket{-}\bra{-}\hat{D}_X(-\alpha(t))\hat{D}_Y(-\beta(t))$ where $\hat{D}_X(\alpha(t))$ and $\hat{D}_Y(\beta(t))$ are the displacement operators defined as $e^{\alpha(t)\hat{a}^\dag_X-\alpha(t)^*\hat{a}_X}$ and $e^{\beta(t)\hat{a}^\dag_Y-\beta(t)^*\hat{a}_Y}$. Applying this to the initial state of the ion, $\ket{\psi(t=0)}=\ket{\downarrow}\ket{0}\ket{0}$, after sideband cooling and qubit initialization, we get the following wave function which exhibits spin-motion entanglement in both motional modes:
\begin{equation}
    \ket{\psi(t)}=\frac{1}{\sqrt{2}}(\ket{+}\ket{\alpha(t)}\ket{\beta(t)}+
    \ket{-}\ket{-\alpha(t)}\ket{-\beta(t)})
\label{eq2}
\end{equation}
\par The time evolution of the spin state for various ratios of detunings to the X and Y modes, $R=\delta_X/\delta_Y$, is presented in \ref{fig:sdf}(a)-(d). The dashed lines are fits to the following equation \cite{haljan_spin-dependent_2005}
\begin{equation}
    P_\uparrow\left(t\right)=\frac{1}{2}\left(1-e^{-\left({\bar{n}}_X+\frac{1}{2}\right)\left|2\alpha\left(t\right)\right|^2-\left({\bar{n}}_Y+\frac{1}{2}\right)\left|2\beta\left(t\right)\right|^2}e^{-t/\tau}\right)
\label{eq3}
\end{equation}
where $\tau$ is an empirical decoherence rate and ${\bar{n}}_X$ and ${\bar{n}}_Y$ are mean phonon numbers of the X and Y modes, which in our system are $\simeq0.2$ and $\simeq0.1$, respectively. In each phase space, the wave packets periodically move in a circular trajectory whose period is defined by the inverse of the detuning of the bichromatic beam. When only one of the motional modes return to the origin in the phase space, the spin states only partially interfere and the measured spin state deviates from its original state, $\ket{\downarrow}$. When the wave packets return to the origin in both phase spaces at the same time, the spin state fully returns to the initial state \cite{haljan_spin-dependent_2005}.

\subsection{Generation of entangled coherent state and observation of phonon number parity modulation}
\begin{figure*}[t]
  \centering
  \includegraphics[width= \textwidth]{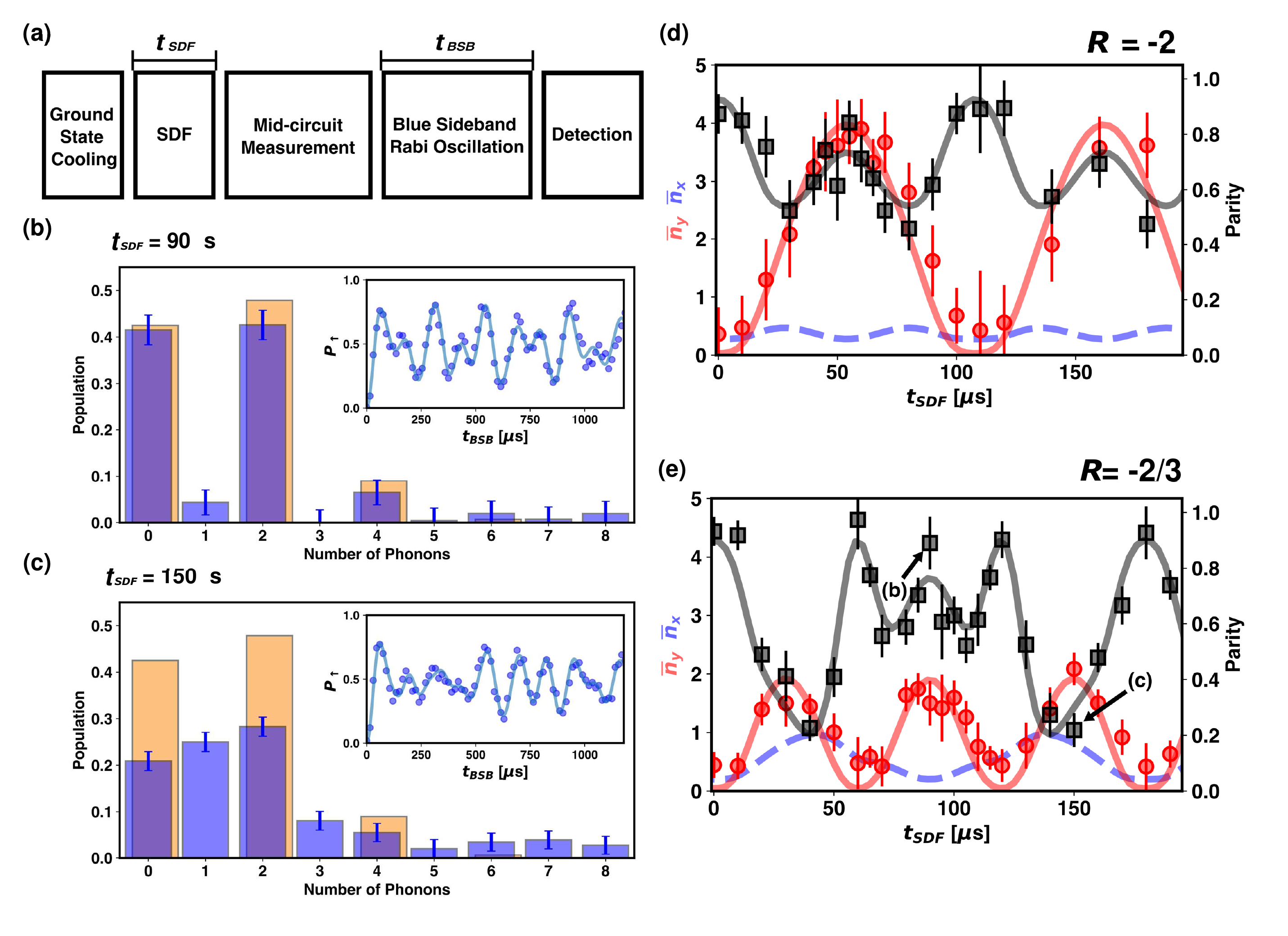}
    \caption{\textbf{Generation of entangled cat state and measurement of phonon state distribution.} (a) Experimental sequence used to generate entangled coherent state and observe the modulation of its parity. (b) A representative plot for phonon distribution of the Y mode with $R=-2/3$ when the X mode is disentangled and (c) entangled. Orange bars are the theoretically expected phonon population for a cat state and blue bars are population extracted by fitting the blue sideband Rabi oscillation, which is presented in the insets. Solid curves in the insets are fits to the blue sideband Rabi oscillation model. (d), (e) Evolution of parity and mean phonon numbers as a functions of $t_{SDF}$ for $R=-2$ and $R=-2/3$, respectively. In (d), the maximum magnitude of the displacement in the Y phase space is $|\beta|=\sqrt{\overline{n}_Y}\simeq2$ and for the X phase space, $|\alpha|=\sqrt{\overline{n}_X}\simeq0.7$. In (e), $|\beta|\simeq1.5$ and $|\alpha|\simeq1.0$ at maximum. Black line is a fit to phonon number parity of the entangled coherent state. Solid red line is the mean phonon number in the Y mode derived from the Rabi frequency and temperatures of each mode obtained from the phonon number parity fitting. Dahsed blue line is the mean phonon number of the X mode calculated the same way. All the error bars in this figure represent standard errors of fitted parameters.}
\label{fig:ecs}
\end{figure*}

The tripartite entangled state of spin and the two motions can be transformed into an entangled coherent state (ECS) of the two motional degrees of freedom by projecting the spin state with mid-circuit measurement. Modifying this sequence to displace only a single motional mode will produce a single-mode cat state of motion, as experimentally shown by Kienzler \textit{et al} (see the Supplementary Material) \cite{kienzler_observation_2016}. We start the experimental sequence by cooling the ion to the ground state with sideband cooling pulses. Then we apply the two-mode SDF, which acts on both the X and Y motions simultaneously, for a duration of $t_{SDF}$. In the following step, the ion is irradiated with a near-resonant 369.5-nm laser beam that serves as the detection beam. It is turned on for 500 $\mu$s, and the scattered photons are collected by a photomultiplier tube. We then drive a blue sideband Rabi oscillation on the Y mode for varying amounts of time and measure the spin state of the ion. This sequence is shown in \ref{fig:ecs}(a). We post-select the wave function with $\ket{\downarrow}$ spin state which is heralded by the detection of less than two photons during the mid-circuit detection phase. 

This results in the following wave function:
\begin{equation}
    \ket{\psi_{ECS}(t)}=\ket{\downarrow}\frac{\ket{\alpha(t)}\ket{\beta(t)}+\ket{-\alpha(t)}\ket{-\beta(t)}}{\sqrt{2+2e^{-2(|\alpha(t)|^2+|\beta(t)|^2)}}}
\label{eq4}
\end{equation}

\par The complimentary data sets which have more than or equal to two photons detected correspond to the entangled coherent state with the opposite phase $\ket{\alpha(t)}\ket{\beta(t)}-\ket{-\alpha(t)}\ket{-\beta(t)}$, but we choose not to use them because photon scattering affects coherence of motional states via recoil \cite{kienzler_observation_2016}. The blue sideband Rabi oscillation is then fitted to the following model to retrieve phonon number distribution of the Y mode \cite{meekhof_generation_1996, turchette_decoherence_2000, kienzler_observation_2016}
\begin{equation}
    P_\uparrow\left(t_{BSB}\right)=\sum_{n=0}^{N}{\frac{p_{Y,n}(t)}{2}\left(1-\cos\left(\Omega_{n+1,n}t_{BSB}\right)e^{-t_{BSB}/\tau}\right)}
\label{eq5}
\end{equation}
where $p_{Y,n}(t)$ is the population for $n$-phonon state in Y-motion after applying the SDF for $t$, and $N$ is the maximum phonon number we consider, $\Omega_{n+1,n}$ is the first order blue sideband Rabi frequency for $n$-phonon state, and $\tau$ is the coherence time. When $\alpha=0$, the wave function in \ref{eq4} is reduced to a single-mode cat state of the Y mode $(\ket{\psi_{Y}(t)}=\ket{\downarrow}(\ket{\beta(t)}+\ket{-\beta(t)})/\sqrt{2+2e^{-2|\beta(t)|^2}})$ and the phonon population is expected to be only in the even number states. However, for a non-zero $\alpha$, the interference between the two coherent states with opposite phases in the Y phase space, $\ket{\beta}_Y$ and $\ket{-\beta}_Y$, is suppressed by the motion in the X-axis, $\ket{\alpha}_X$ and $\ket{-\alpha}_X$. Consequently, the parity of the Y mode population is modulated as the size of the displacement in the X mode changes. The resulting time evolution of the phonon population distribution is as follows.

\begin{equation}
\begin{split}
    p_{Y,n}(t)=Tr(\{\ket{\downarrow}\bra{\downarrow}\otimes\hat{I}_X\otimes\ket{n}\bra{n}\}\ket{\psi_{ECS}}\bra{\psi_{ECS}})\\
    =\frac{e^{-|\beta\left(t\right)|^2}(|\beta\left(t\right)|^{2n}/n!)}{1+e^{-2\left(\left|\alpha\left(t\right)\right|^2+\left|\beta\left(t\right)\right|^2\right)}}(1+\left(-1\right)^ne^{-2\left|\alpha\left(t\right)\right|^2})
\end{split}
\label{eq6}
\end{equation}
where $\bra{n}$ is the $n$-th number state of the Y mode. \ref{eq6} results in a modulation of phonon number parity defined as $\Pi\left(t\right)=\sum_{n}{\left(-1\right)^np_{Y,n}(t)}$, because the interference between the odd number states is suppressed by entanglement with the X mode. We include the effect of imperfect sideband cooling in the model, which leaves some population in the $1$-phonon state (see Methods). We demonstrate the generation of entangled coherent states at various values of $R=\delta_X/\delta_Y$. \ref{fig:ecs}(d) corresponds to $R=-2$, where the ion is periodically displaced in the X axis at a frequency which is twice of the frequency of the periodic displacement in the Y axis. Therefore, according to \ref{eq6}, the parity of the phonon distribution of the Y motion is expected to be modulated at half the period of its periodic displacement. 
\par We repeat the same experiment with $R=-2/3$. In this case, the parity modulation pattern is expected to span three periods of the Y displacement as shown in \ref{fig:ecs}(e). The observed variation in phonon number parity is in good agreement with the theoretical model, and is a direct consequence of the entanglement of the two motional modes. \ref{fig:ecs}(b) and (c) are the two representative phonon distributions. The Y mode displacement is maximum for both, but the phonon number parity is $0.89\pm0.09$ for \ref{fig:ecs}(b) and $0.22\pm0.06$ for \ref{fig:ecs}(c). Also, \ref{fig:ecs}(c) shows a clear deviation from the single-mode cat state phonon distribution with a significant population in the $\ket{1}_Y$ and $\ket{3}_Y$ states. In \ref{fig:ecs}(d) and (e), we also plotted the time evolution of the mean phonon numbers of the Y mode, which approximately corresponds to the square of the absolute value of the displacement in the Y mode phase space. The theoretical curves for the mean phonon numbers of the X and Y modes are calculated by using the Rabi frequency and $1$-phonon state population of each mode inferred by fitting the phonon number parity data.

\begin{figure*}[t]
  \centering
  \includegraphics[width= \textwidth]{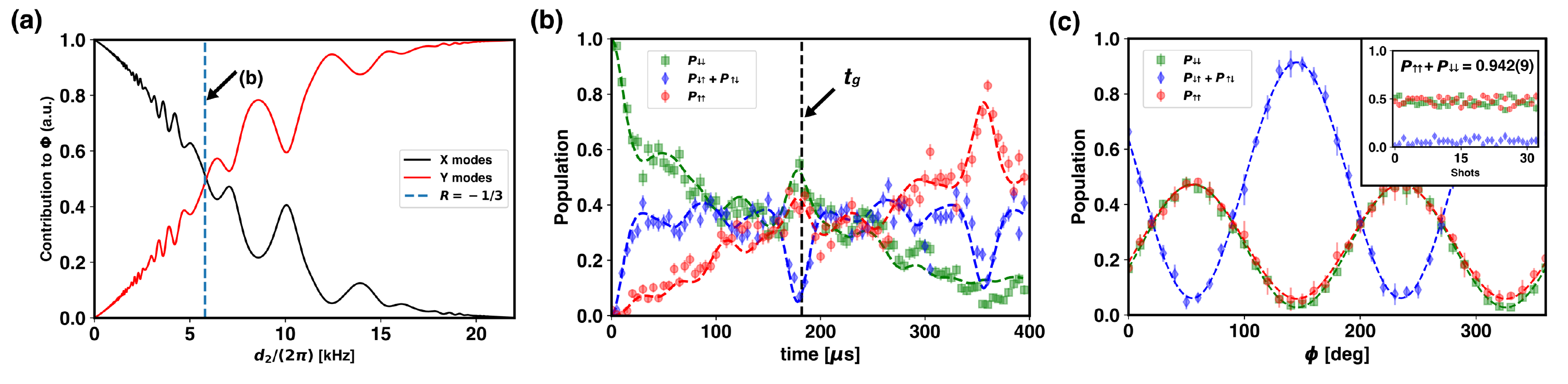}
    \caption{\textbf{Characterization of Mølmer–Sørensen gate with two-dimensional motion.} (a) Normalized contributions of each mode for the geometric phase needed for the generation of the Bell state, $1/\sqrt{2}(\ket{\downdownarrows}+\ket{\upuparrows})$. $d_{2}$ is the detuning from the $X_{cm}$ mode. At the detuning indicated by a vertical dashed line, the detuning ratio is $R=-1/3$ and the X and Y mode contribute almost equally. (b) Time evolution of the spin states of a two-qubit system under two-dimensional Mølmer–Sørensen interaction when $R=-1/3$. The optimal gate time $t_g=182\ \mu s$. Error bars represent quantum projection noise. (c) Qubit state population oscillation as a function of the phase of the $\pi/2$-pulse. Qubit state parity(not shown in the plot), $\Pi(\phi)=P_{\uparrow\uparrow}(\phi)+P_{\downarrow\downarrow}(\phi)-(P_{\downarrow\uparrow}(\phi)+P_{\uparrow\downarrow}(\phi))$, oscillates with an amplitude of $\Pi_a=0.852\pm0.007$. Error bars are the standard deviation calculated from five iterations of the same experiment. The average population of the even states at $t_g$, $P_{\uparrow\uparrow}+P_{\downarrow\downarrow}$, is $0.942\pm0.009$ as shown in the inset. The parity oscillation and even state population yield a gate fidelity of $0.897\pm0.006$.}
\label{fig:ms}
\end{figure*}

\subsection{Mølmer–Sørensen gate with two-dimensional motion}   

Next, we trap two ions in a linear chain and investigate how the two-dimensional coherent motion can be utilized in an ion chain by realizing Mølmer–Sørensen interaction \cite{sorensen_entanglement_2000, benhelm_towards_2008, lee_phase_2005} involving modes from multiple principal axes. We first observe the time evolution of the two-qubit states under two-dimensional Mølmer–Sørensen interaction, with the Rabi frequency and detuning calibrated to generate a Bell state ($1/\sqrt{2}(\ket{\upuparrows}+\ket{\downdownarrows})$). Then we measure the fidelity of the resulting state by observing qubit state parity oscillation, which is implemented by applying a $\pi/2$-pulse that acts on both qubits and scanning its phase, $\phi$ \cite{benhelm_towards_2008}. The gate fidelity is estimated at $R=-1/3$, which in this context is defined as the ratio between the detunings to the X-cm mode and the Y-tilt mode, $\delta_{X_{cm}}/{\delta_Y}_{tilt}$. The measured gate fidelity is $89.7\pm0.6$ \% which is comparable to our single axis Mølmer–Sørensen gate fidelity, $93.2\pm0.6$ \%. This indicates that Mølmer–Sørensen interaction can be expanded into multiple dimensions naturally. Also, the gate Rabi frequency is reduced compared to the single axis case, because more phase spaces contribute to the geometric phase ($\Phi_n(t)={\eta_{n1}\eta_{n2}/(2d_n)^2\left(d_nt-\sin\left(d_nt\right)\right)\Omega_0^2}$ where $\eta_{n1}$ and $\eta_{n2}$ are the Lamb-Dicke factors of each ion for the $n$-th mode, $d_n$ is the detuning to $n$-th mode and $\Omega_0$ is the Rabi frequency), as can be seen in \ref{fig:ms}(a). This effect is most pronounced at $R=-1/3$ where the geometric phase contribution is similar for both axes, thus the required Rabi frequency is reduced by a factor of $\simeq 1/\sqrt{2}$. Here, the Rabi frequency needed to generate an equal superposition of $\ket{\downdownarrows}$ and $\ket{\upuparrows}$ using both axes is $2\pi\times86.1$ kHz, which is in agreement to the experimentally calibrated value of $2\pi\times81.3\pm0.6$ kHz. This is $28.3 \%$ lower compared to the Rabi frequency required using only the X axis, and $30.1 \%$ lower compared to using only the Y axis, assuming the same gate time and gate detuning.

\section{Discussion}
In this work, we have demonstrated the generation of entangled coherent states with two-dimensional motion of a trapped ion. Our scheme uses the near-degeneracy of the transverse modes of a linear Paul trap to excite the two motional modes of a single ion simultaneously with detuned SDF, and does not require second order interactions or two-phonon interactions as proposed in \cite{gerry_generation_1997, zou_generation_2001}, which is advantageous in terms of the strength of the interaction. We observed a periodic modulation in the single-mode phonon number parity which is a direct consequence of the entanglement between the two phonon modes. The loss of parity information is analogous to the decoherence of the spin state when only the spin state is directly measured in a spin-motion entangled system \cite{haljan_spin-dependent_2005}. We have also shown that Mølmer–Sørensen interaction with multiple ions can be easily realized with a lower laser power using a two-dimensional spin-dependent force.

\par The size of the phase space displacements produced in our experiment is mainly limited by the large Rabi frequency required to generate a strong SDF in both X and Y modes. For a SDF with non-zero detuning, the maximum displacement is limited because the phase space trajectory forms a circle with radius proportional to the inverse of the detuning. Thus, the maximum displacement can be increased by making the trap more isotropic in the transverse directions. Alternatively, one can apply a SDF resonant to both the X and Y modes, which can be realized by a tetrachromatic laser beam \cite{solano_entangled_2002}. In this case the size of the displacement will increase linearly with $t_{SDF}$ and the coupling of the laser to each mode. Another limiting factor is the difficulty of characterizing motional states with large displacements via blue sideband Rabi oscillation. States with larger displacement magnitude are harder to probe because the coherence time of a cat state scales inversely with the square of magnitude of displacement \cite{turchette_decoherence_2000}.
\par The creation of an entangled coherent state with opposite phase, $\ket{\downarrow}(\ket{\alpha}\ket{\beta}-\ket{-\alpha}\ket{-\beta})$, is possible with a  $\pi$-pulse phase-locked to the SDF preceding the mid-circuit measurement \cite{kienzler_observation_2016}. 
\par Using our scheme, up to $3N$ modes can be entangled for an N-ion chain when all the principal axes of the trapping potential are utilized. Especially, the generation of a tripartite entangled coherent state of the X, Y and Z modes, combined with a beam splitter interaction between the phonon modes \cite{zhang_noon_2018, gan_hybrid_2020, toyoda_hongoumandel_2015}, will enable the quantum teleportation protocol in Ref. \cite{van_enk_entangled_2001} with a single trapped ion.

\par There have been proposals and experiments of a Ramsey-type matter-wave rotation sensor \cite{campbell_rotation_2017, shinjo_three-dimensional_2021}, Rabi-type sensor \cite{martinez-garaot_interferometer_2018} and Rashba and Dresselhaus-type spin-orbit coupling for quantum simulation of topological insulators and Majorana fermions in which a single ion is coherently manipulated in two or more orthogonal spatial modes \cite{rossetti_trapped-ion_2016}. The coherent control of two-dimensional motion of a trapped ion demonstrated in this work can be applied to realize such experiments. Lastly, utilization of quantum motion in multiple axes for the realization of entangling gates can reduce the experimental overhead required to suppress interactions in multiple directions often employed in trapped ion quantum computing setups, such as asymmetric trap geometry \cite{madsen_planar_2004} and trap RF voltage offset \cite{yum_optical_2020}.

\section{Methods}
\subsection{Instruments}
The pulsed laser beams are provided by a 355-nm mode-locked laser (Coherent, Paladin Compact 355-4000). Its repetition rate fluctuates around 120.1 MHz due to the thermal and acoustic perturbations in the laser cavity. The repetition rate is monitored by an ultrafast photodetector (Alphalas, UPD-30-VSG-P). The drift is compensated by an AOM in the path of the $R_V$ beam whose RF frequency is updated at a rate of ~50 kHz by a field-programmable-gate-array (Digilent, Arty-S7) running a custom PID program \cite{mount_scalable_2016}. The RF trap voltage is sampled by a capacitive divider and rectified by a diode circuit \cite{park_new_2021}. The rectified voltage is fed to a high-speed PID controller (New Focus, LB1005-S), which controls the output power of the RF source to stabilize the trap RF voltage \cite{johnson_active_2016}.
\subsection{Experimental protocol}
To drive the two-dimensional motion of a single trapped ion along a closed trajectory, we first optimize the Raman detuning to a frequency where the phase space trajectories for the X and Y axis can be closed simultaneously. The protocol is as follows: (1) The ion is ground-state cooled by sequentially applying sideband cooling pulses to the X and Y modes. (2) The spin state of the ion is initialized to $\ket{\downarrow}$ via optical pumping. (3) With the probe time set to $T=2\pi\times N/(\omega_Y-\omega_X)$ where N is an integer, we scan the Raman detuning between $\omega_Y$ and $\omega_X$ and look for frequencies where the measured spin state is close to $\ket{\downarrow}$, which indicates the simultaneous disentanglement of the spin from the motional states in the X and Y axes. (4) To balance the red and blue sideband transitions and calibrate out the differential Strak shift, the RF power and frequency for the transitions are fine-tuned individually to minimize the $\ket{\uparrow}$ state population.
\par In the entangled coherent state experiment, we limit the Rabi frequency of the blue sideband transition used to probe the phonon distribution of the Y mode to about $5$ kHz, so as not to excite the blue sideband transition of the X mode. At this value, the expected maximum amplitude of the X mode blue sideband Rabi oscillation is $0.7 \%$, thus we did not include the excitation of the X mode in the phonon distribution analysis.

\par The relatively low frequency of the blue sideband Rabi oscillation means that even a small drift in the secular frequency of the trap can affect the phonon state estimation results. The trap RF power is stabilized by a PID loop, but it drifts slowly due to the temperature changes in the components of the PID circuit at a rate of 2 kHz/hr in the worst case. Thus, we interleave a blue sideband Ramsey spectroscopy experiment with the main experiment for every data point in \ref{fig:ecs}(d) and (e), and monitored the change of secular frequency throughout data collection process. Data collection for each point in the figures takes about 5 minutes and for all the points in each plot about 2 hours. We stop the experiment if the secular frequency of the Y mode changed from the calibrated value by more than 300 Hz. We recalibrate the frequencies for the blue sideband transition and the spin-dependent force, and then resume the experiment. With a Rabi frequency of 5 kHz and detuning of 300\ Hz, the amplitude of the blue sideband Rabi oscillation decreases by $0.4 \%$ and Rabi frequency increases by $0.9 \%$, which are negligible for the purposes of our experiment. Also, we note that in the analysis of the blue sideband Rabi oscillation, we use the exact form of $\Omega_{n+1,n}=\Omega_{0,0}\bra{n+1}e^{i\eta_Y(\hat{a}^{\dag}_Y+\hat{a}_Y)}\ket{n}=\Omega_{0,0}exp{\left(-\eta_Y^2/2\right)}\eta_Y/\sqrt{n+1}{\ L}_n^1(\eta_Y^2)$ where $\Omega_{0,0}$ is the carrier transition Rabi frequency with zero phonons, $\eta$ is the Lamb-Dicke factor, and $L_n^1$ is the generalized Laguerre polynomial of \textit{n}-th order \cite{leibfried_quantum_2003}, since in our data the maximum value of $\eta_Y\sqrt{2{\bar{n}}_Y+1}$ is about 0.33 where the Lamb-Dicke approximation becomes inaccurate.

\par Also, to eliminate the possibility of the slow drift during the experiment affecting the observed pattern of parity modulation, we conducted the experiment in randomized orders of $t_{SDF}$. The full randomized sequences of $t_{SDF}$ used for the data sets in \ref{fig:ecs}(d) and (e) are available in the Supplementary Material.

\subsection{Effect of finite temperature on phonon number parity}
The measured maximum value of parity shown in \ref{fig:ecs} does not reach unity because of imperfect sideband cooling, which in our setup typically results in $\overline{n}_X\simeq0.2$ and $\overline{n}_Y\simeq0.05$ . This finite temperature effect is modelled by considering a mixed motional state with a population of $p_{X,1}$ and $p_{Y,1}$ in the first excited state of each mode and the rest in the motional ground states. We include the following three initial states in the model. (i) $\ket{1}_X\ket{0}_Y$ with probability $p_{X,1}\left(1-p_{Y,1}\right)$, (ii) $\ket{0}_X\ket{1}_Y$ with probability $\left(1-p_{X,1}\right)p_{Y,1}$, and (iii) $\ket{0}_X\ket{0}_Y$ with probability $\left(1-p_{X,1}\right)\left(1-p_{Y,1}\right)$. The $\ket{1}_X\ket{1}_Y$ state is not considered since its probability is negligible. When the motional state is the \textit{n}-th excited state, the effect of the displacement operator and the resulting phonon distribution can be calculated using number state representations of the displacement operator, $d_{mn}^\alpha=\bra{m}\hat{D}(\alpha)\ket{n}$. We employed the results of \cite{cahill_ordered_1969} to calculate $d_{mn}^\alpha$.
\par For (i), the modified phonon distribution of the Y mode is as follows
\begin{equation*}
\begin{split}
    p_{Y,n}\left(t\right)=\frac{1}{\left(1+e^{-2\left(\left|\alpha\left(t\right)\right|^2+\left|\beta\left(t\right)\right|^2\right)}\right)}e^{-\left|\beta\left(t\right)\right|^2}\frac{\left|\beta\left(t\right)\right|^{2n}}
    {n!}\\
    \times\left(1+\left(-1\right)^n\frac{\left(d_{11}^{-2\alpha}+d_{11}^{2\alpha}\right)}{2}\right)
\end{split}
\end{equation*}

\par For (ii), 
\begin{equation*}
\begin{split}
    p_{Y,n}\left(t\right)=\frac{1}{2\left(1+e^{-2\left(\left|\alpha\left(t\right)\right|^2+\left|\beta\left(t\right)\right|^2\right)}\right)}\\
    \times(\left|d_{n1}^\beta\right|^2+\left|d_{n1}^{-\beta}\right|^2+\left(\left(d_{n1}^\beta\right)^\ast d_{n1}^{-\beta}+d_{n1}^\beta{(d_{n1}^{-\beta})}^\ast\right)e^{-2\left|\alpha\right|^2})
\end{split}
\end{equation*}.
\par The weighted sum of the phonon distributions corresponding to the above three cases were used to fit the parity modulation data and extract the Rabi frequency, $p_{X,1}$ and $p_{Y,1}$.

\section{Data Availability}
The data supporting the findings in this study are available from the corresponding author upon reasonable request.

\section{Acknowledgments}
This work was supported by Institute for Information \& communications Technology Planning \& Evaluation (IITP) grant (No.2022-0-01040), the National Research Foundation of Korea (NRF) grant (No. 2020R1A2C3005689, No. 2022M3E4A1083521), and the Samsung Research Funding \& Incubation Center of Samsung Electronics (No. SRFC-IT1901-09).

\section{Author Contributions}
H.J. designed, carried out and analyzed the experiments. J.Y.K. carried out the experiments and implemented the experimental control system and the feed-forward system. H.J., J.Y.K., J.E.K., W.H.C., and K.K. contributed to the setup of the experimental apparatus. H.J. wrote and edited the manuscript. T.K. supervised the experiment. All authors participated in the discussion of the experimental results.

\bibliography{ref}

\end{document}


\title{Supplementary Material\\ $\ $ \\Experimental Realization of Entangled Coherent States in Two-dimensional Harmonic Oscillators of a Trapped Ion}
\author{Honggi Jeon}
\affiliation{Department of Physics and Astronomy, Seoul National University, Seoul 08826, Republic of Korea}
\affiliation{Automation and System Research Institute, Seoul National University, Seoul 08826, Republic of Korea}
\author{Jiyong Kang}
\affiliation{Automation and System Research Institute, Seoul National University, Seoul 08826, Republic of Korea}
\affiliation{Department of Computer Science and Engineering, Seoul National University, Seoul 08826, Republic of Korea}
\author{Jaeun Kim}
\affiliation{Automation and System Research Institute, Seoul National University, Seoul 08826, Republic of Korea}
\affiliation{Department of Computer Science and Engineering, Seoul National University, Seoul 08826, Republic of Korea}
\author{Wonhyeong Choi}
\affiliation{Automation and System Research Institute, Seoul National University, Seoul 08826, Republic of Korea}
\affiliation{Department of Computer Science and Engineering, Seoul National University, Seoul 08826, Republic of Korea}
\affiliation{Inter-university Semiconductor Research Center, Seoul National University, Seoul 08826, Republic of Korea}
\author{Kyunghye Kim}
\affiliation{Automation and System Research Institute, Seoul National University, Seoul 08826, Republic of Korea}
\affiliation{Department of Computer Science and Engineering, Seoul National University, Seoul 08826, Republic of Korea}
\author{Taehyun Kim}
\email{taehyun@snu.ac.kr}
\affiliation{Automation and System Research Institute, Seoul National University, Seoul 08826, Republic of Korea}
\affiliation{Department of Computer Science and Engineering, Seoul National University, Seoul 08826, Republic of Korea}
\affiliation{Inter-university Semiconductor Research Center, Seoul National University, Seoul 08826, Republic of Korea}
\affiliation{Institute of Computer Technology, Seoul National University, Seoul 08826, Republic of Korea}
\affiliation{Institute of Applied Physics, Seoul National University, Seoul 08826, Republic of Korea}

\maketitle

\section{Phonon Distribution Extraction Procedure}
We used a simple exponential decay model of the following form to analyze the blue-sideband Rabi oscillation results shown in Fig. 3 of the main text \cite{meekhof_generation_1996, turchette_decoherence_2000, kienzler_observation_2016}.
\begin{equation}
    P_\uparrow\left(t_{BSB}\right)=\sum_{n=0}^{N}{\frac{p_{Y,n}}{2}\left(1-cos\left(\Omega_{n+1,n}t_{BSB}\right)e^{-t_{BSB}/\tau}\right)}
\label{eq1}
\end{equation}
where $P_\uparrow\left(t_{BSB}\right)$ is the probability for the qubit state to be $\ket{\uparrow}$ at $t_{BSB}$, $p_{Y,n}$ is the phonon-n state population of the Y mode, $\Omega_{n+1,n}$ is the blue sideband Rabi frequency, and $\tau$ is the coherence time. The phonon population distribution is inferred by fitting the blue sideband Rabi oscillation data to \ref{eq1}. When fitting the data, we set the maximum phonon number in the model, $N$, to 8 for $R=-2/3$ and to 12 for $R=-2$ to prevent overfitting. For all data points, the initial guess for the phonon population distribution is set to that of an even cat state. Also, we set the maximum amplitude of the Rabi oscillation to $0.97$, because the detection error for the $\ket{\uparrow}$ state using the threshold method is $3 \%$ in our system, although changing it to 1.0 changes the results in Fig. 3 of the main text negligibly. 
\par
\begin{figure*}[!h]
  \centering
  \includegraphics[width= 0.6\textwidth]{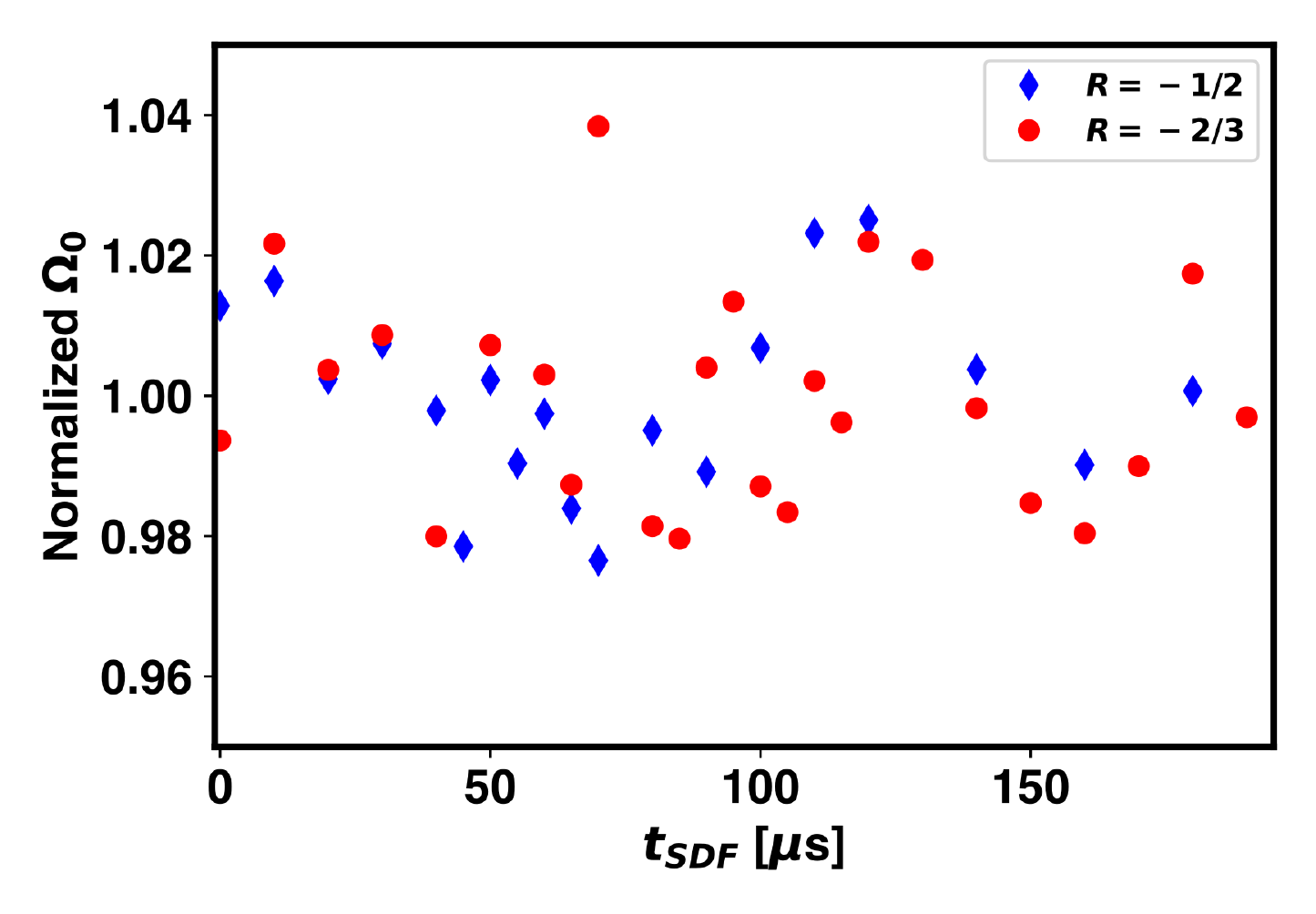}
    \caption{Values of the fitted $\Omega_0$ for the two data sets in Fig. 3 in the main text. The fitted Rabi frequencies vary less than 5 \%, as expected.}
\label{fig:fitted_rabi}
\end{figure*}

 Lastly, the initial guess for the carrier Rabi frequency, $\Omega_0$, for the least square algorithm that extracts phonon number distribution from blue sideband Rabi oscillation had to be chosen carefully because assuming an incorrect Rabi frequency will give you an inaccurate distribution. We consider the carrier Rabi frequency because the fitting function converts it into blue sideband Rabi frequency by multiplying it with relevant factors including the Y axis Lamb-Dicke factor, $\eta_Y$. We chose the initial Rabi frequency that gives the highest phonon population for the $\ket{0}_Y$ state for the $t_{SDF} = 0$ $\mu$s data. Throughout a single data set, the fitted Rabi frequency varied by less than $5 \%$, as shown in \ref{fig:fitted_rabi}. 

\clearpage

The fitted values and standard errors for the parity modulation in Fig. 3 of the main text are as follows:

\begin{table}[!ht]
    \centering
    \begin{tabular}{|c|c|c|}
    \hline
         ~ & $R=-1/2$ & $R=-2/3$ \\ \hline
        $\Omega_{SDF}/(2\pi)$ [kHz] & $167.683\pm17.157$ & $212.600\pm6.389$ \\ \hline
        $p_{X,1}$ & $0.287\pm0.073$ & $0.213\pm0.074$ \\ \hline
        $p_{Y,1}$ & $0.043\pm0.024$ & $0.056\pm0.015$ \\ \hline
    \end{tabular}
    \caption{Results of parity fitting obtained from Fig 3. (d) and (e) of the main text}
\end{table}

\clearpage

\section{Sequences of $t_{SDF}$ used for entangled coherent state experiments}
The data collection process for each data point in Fig. 3(d) and (e) took about five minutes, thus the total data collection time for each of the two data sets spanned about two hours. To prevent the slow drift of the trap frequency affecting the observed variation in phonon number parity, we randomized the data taking sequence as presented below:

\begin{table}[!ht]
    \centering
    \begin{tabular}{|c|c|c|c|c|c|c|c|c|c|c|c|c|c|c|c|c|c|c|c|c|c|c|c|}
    \hline
        Sequence & 1 & 2 & 3 & 4 & 5 & 6 & 7 & 8 & 9 & 10 & 11 & 12 & 13 & 14 & 15 & 16 & 17 & 18 & 19 & 20 & 21 & 22 & 23 \\ \hline
        $R=-2/3$ & ~ & ~ & ~ & ~ & ~ & ~ & ~ & ~ & ~ & ~ & ~ & ~ & ~ & ~ & ~ & ~ & ~ & ~ & ~ & ~ & ~ & ~ & ~ \\ 
        $t_{SDF} [\mu s]$ & 110 & 0 & 120 & 190 & 20 & 65 & 60 & 115 & 10 & 80 & 140 & 100 & 50 & 160 & 70 & 150 & 130 & 105 & 90 & 40 & 30 & 170 & 180 \\ \hline
        $R=-2$ & ~ & ~ & ~ & ~ & ~ & ~ & ~ & ~ & ~ & ~ & ~ & ~ & ~ & ~ & ~ & ~ & ~ & ~ & ~ & ~ & ~ & ~ & ~ \\ 
        $t_{SDF} [\mu s]$ & 65 & 110 & 90 & 70 & 100 & 55 & 60 & 40 & 30 & 80 & 0 & 120 & 10 & 50 & 45 & 20 & 160 & 180 & 140 & ~ & ~ & ~ & ~ \\ \hline
    \end{tabular}
    \caption{Values and randomized sequences of $t_{SDF}$ used to obtain experimental data in Fig. 3 (d) and (e) of the main text}
\end{table}
\clearpage

\section{Single-mode cat state}
\begin{figure*}[!h]
  \centering
  \includegraphics[width= \textwidth]{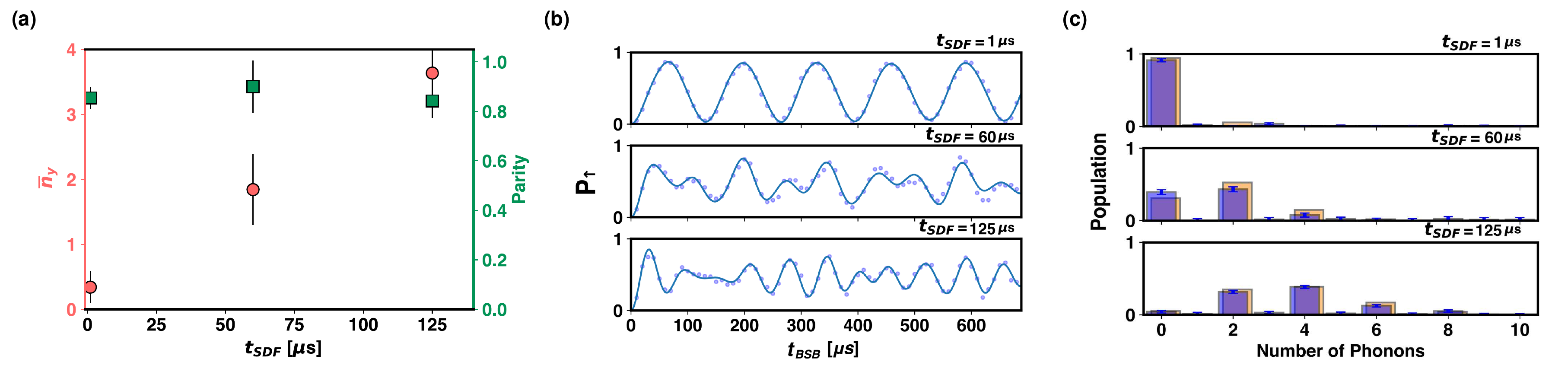}
    \caption{\textbf{Single mode cat state analysis}. (a) Phonon number fitting results for a single-mode cat state in which only the Y motion is excited. The parity of the phonon number state remains high while the mean phonon number increases. (b) Blue sideband Rabi oscillation of the Y mode at various $t_{SDF}$. Solid lines are fits to phonon distribution model with a maximum phonon number of 10. (c) Phonon number distribution for each $t_{SDF}$ extracted by fitting (b) to \ref{eq1}. Blue bars are the experimentally measured phonon population and orange bars are the theoretically expected phonon population for a single-mode cat state for the measured mean phonon number. the extracted population matches that of a single-mode even cat state where odd number populations are suppressed. All error bars in this figure indicate standard errors of fitted parameters.}
\label{fig:1dcat}
\end{figure*}

To ensure that our system and analysis scheme can accurately generate and correctly characterize non-classical states, we first created a single mode cat state $(\ket{\psi_{Y}(t)}=\ket{\downarrow}(\ket{\beta(t)}+\ket{-\beta(t)})/\sqrt{2+2e^{-2|\beta(t)|^2}})$ by driving 1D SDF with the Y mode \cite{kienzler_observation_2016}. The results are presented in \ref{fig:1dcat}.

\section{Population under Mølmer–Sørensen interaction}
The populations of the qubit states in a two-ion chain evolve under Mølmer–Sørensen interaction with four motional states as follows:
\begin{equation*}
    P_{\downarrow\uparrow}(t)=P_{\uparrow\downarrow}(t)=\frac{1}{8}(2-e^{-8\left((\overline{n}_1+\frac{1}{2})|\alpha_1(t)|^2+(\overline{n}_3+\frac{1}{2})|\alpha_3(t)|^2\right)}-e^{-8\left((\overline{n}_2+\frac{1}{2})|\alpha_2(t)|^2+(\overline{n}_4+\frac{1}{2})|\alpha_4(t)|^2\right)})
\label{eq2}
\end{equation*}

\begin{equation*}
\begin{split}
    P_{\downarrow\downarrow}(t)=\frac{1}{8}(2+e^{-8\left((\overline{n}_1+\frac{1}{2})|\alpha_1(t)|^2+(\overline{n}_3+\frac{1}{2})|\alpha_3(t)|^2\right)}+e^{-8\left((\overline{n}_2+\frac{1}{2})|\alpha_2(t)|^2+(\overline{n}_4+\frac{1}{2})|\alpha_4(t)|^2\right)}\\
    +4\cos\left(4\Phi(t)\right)e^{-2((\overline{n}_1+\frac{1}{2})|\alpha_1(t)|^2+(\overline{n}_2+\frac{1}{2})|\alpha_2(t)|^2+(\overline{n}_3+\frac{1}{2})|\alpha_3(t)|^2+(\overline{n}_4+\frac{1}{2})|\alpha_4(t)|^2)})
\label{eq3}
\end{split}
\end{equation*}

\begin{equation*}
\begin{split}
    P_{\uparrow\uparrow}(t)=\frac{1}{8}(2+e^{-8\left((\overline{n}_1+\frac{1}{2})|\alpha_1(t)|^2+(\overline{n}_3+\frac{1}{2})|\alpha_3(t)|^2\right)}+e^{-8\left((\overline{n}_2+\frac{1}{2})|\alpha_2(t)|^2+(\overline{n}_4+\frac{1}{2})|\alpha_4(t)|^2\right)}\\
    -4\cos\left(4\Phi(t)\right)e^{-2((\overline{n}_1+\frac{1}{2})|\alpha_1(t)|^2+(\overline{n}_2+\frac{1}{2})|\alpha_2(t)|^2+(\overline{n}_3+\frac{1}{2})|\alpha_3(t)|^2+(\overline{n}_4+\frac{1}{2})|\alpha_4(t)|^2)})
\label{eq4}
\end{split}
\end{equation*}

\begin{equation*}
    \Phi\left(t\right)=\sum_{n=1}^4{\frac{\eta_{n1}\eta_{n2}}{(2d_n)^2}\left(d_nt-\sin\left(d_nt\right)\right)\Omega_0^2}
\label{eq5}
\end{equation*}
where $n= 1, 2, 3$ and $4$ is the index for the motional modes participating in the interaction corresponding to $X_{tilt}, X_{cm}, Y_{tilt}$ and $Y_{cm}$ mode, respectively. $\alpha_n(t)$ is the phase space displacement of the $n$-th motional mode at time $t$, $\eta_{nk}$ is the Lamb-Dicke factor for the $n$-th mode and the $k$-th ion, $d_n$ is the laser detuning, $\overline{n}_n$ is the mean phonon number, and $\Omega_0$ is the Rabi frequency. The above formulae were derived by following the calculations presented in \cite{manning_quantum_2014}. These equations were used to analyze and derive the results about  Mølmer–Sørensen interaction in the main text.
\bibliographystyle{apsrev4-2}
\bibliography{ref}